# Automatic analysis of Categorical Verbal Fluency for Mild Cognitive Impartment detection: a non-linear language independent approach

K. López-de-Ipiña, U. Martinez-de-Lizarduy, N. Barroso
Universidad del País Vasco/Euskal Herriko Unibertsitatea
{karmele.ipina, unai.martinezdelizarduy}@ehu.eus

Marcos Faundez-Zanuy
Escola Universitaria Politècnica de Mataró (UPF),
Tecnocampus {faundez, sesa}@tecnocampus.cat

M. Ecay-Torres, P. Martinez-Lage, F. Torres
Fundación CITA Alzheimer
{mecay, pmlage, ftorres}@cita-alzheimer.org

*Abstract*—: Alzheimer's disease (AD) is one the main causes of dementia in the world and the patients develop severe disability and sometime full dependence. In previous stages Mild Cognitive Impairment (MCI) produces cognitive loss but not severe enough to interfere with daily life. This work, on selection of biomarkers from speech for the detection of AD, is part of a wide-ranging cross study for the diagnosis of Alzheimer. Specifically in this work a task for detection of MCI has been used. The task analyzes Categorical Verbal Fluency. The automatic classification is carried out by SVM over classical linear features, Castiglioni fractal dimension and Permutation Entropy. Finally the most relevant features are selected by ANOVA test. The promising results are over 50% for MCI.

## I. INTRODUCTION

Alzheimer's disease (AD) is one the main causes of dementia and death in the world and the patients develop severe disability and sometime full dependence. AD is the most common type of dementia among the elderly. It is characterized by progressive and irreversible cognitive deterioration with memory loss and impairments in judgment and language, together with other cognitive deficits and behavioral symptoms. The cognitive deficits and behavioral symptoms are severe enough to limit the ability of an individual to perform everyday life.. An early and accurate diagnosis of AD helps to increase their quality of life and offers the best possibilities of treating the symptoms of the disease. The diagnosis of definite AD requires the demonstration of the typical AD pathological changes at autopsy [1,2].

During social interaction in everyday life, verbal communication is, for human being, one of the most important ways of expression. It is a complex process, which involves a wide range of cognitive abilities. We are able to produce sounds with a message inside. The biosignal hides our desires, ideas and/or emotions, a piece of our being. In that process the brain has to manage knowledge, memory, language and semantic information. Moreover, these biological samples have the advantage of be very easy to record without invasive and expensive equipments. In the case of Alzheimer's Disease (AD), the deterioration of spoken language immediately affects the patient's ability to naturally interact with his or her al environment, and it is usually also accompanied by alterations in emotional response. [1-3]. In this sense, the cost and technology requirements of some of medical proofs (fMRI, PET) make impossible to apply such biomarkers to all patients or require painful tests. Given these problems, non-invasive intelligent techniques of diagnosis may become valuable tools for early detection of diseases to complement diagnosis and/or monitoring its progress.

Mild cognitive impairment produces cognitive loss but not severe enough to interfere with daily life or independent abilities. These changes not affect daily life and a person with MCI usually don´t have an appropriate diagnosis. MCI could increase the risk of developing Alzheimer's or another dementia and currently its detection is one of the challenges of medical specialists [4]. In addition, some references show that non-specialists and even familiars are not able to identify exactly early AD as well as Mild Cognitive Impairment (MCI) [5,6]. Thus non-technical staff in the habitual environment of the patient could use these methodologies, which include speech and voice analysis, without altering or blocking the patients' abilities, as the spontaneous speech involved in these techniques is not perceived as a stressful test by the patient. Moreover, these techniques are very low-cost and do not require extensive infrastructure or the availability of medical equipment. They are thus capable of yielding information easily, quickly, and inexpensively [7,8]. In this sense it is essential the search for clinically useful screening and diagnostic Tools.

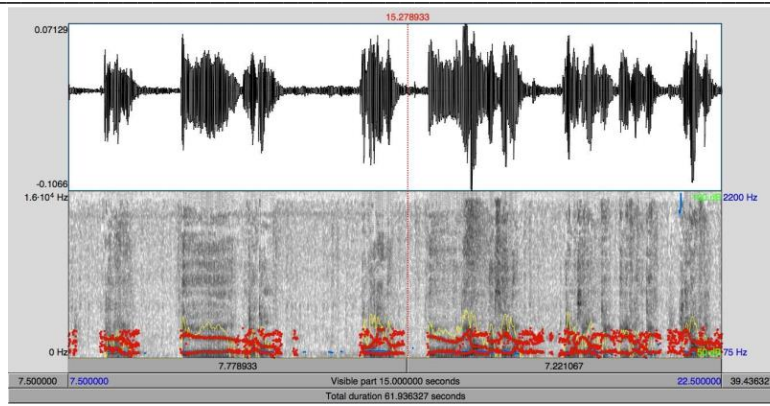

**Figure 1.** Categorical Verbal Fluency (CVF) task uttered by an individual of the control group

The main goal of this project is the development of an automated analysis of categorical verbal fluency (CVF) through speech therapy techniques that allow obtaining these specific analyses quickly and reliably. There are several works as references during the last years. In this paper we focus our research on the integration of the language independent methodologies to detect AD in speech, to a task of Categorical Verbal Fluency (CVF) [1,9]. Specifically the so-called animals naming task.

The materials are described in section II. Section III presents the used methods. The results and discussion are described in section IV finally concluding remarks are included in section V.

## II. MATERIALS

### A. Gipuzkoa-Alzheimer Project (PGA)

The sample consists of 187 healthy people and 38 with MCI belonging to the cohort of Gipuzkoa-Alzheimer Project (PGA) of the CITA-Alzheimer Foundation. The PGA is a longitudinal study focusing on the search for biomarkers of function, brain structure and state of health and risk factors for early diagnosis of Alzheimer's disease. The main objective of PGA is to characterize Alzheimer's disease and pre-clinical prodrome. Inclusion criteria based on the characteristics of each group are listed below.

*1) Control group*

- Men and women of middle age or elderly, between 39 and 79 years (inclusive) to sign the informed consent to participate in the PGA.
- Lack of memory complaints.
- Global cognitive function preserved with have the highest score of 24 in the Mini Mental State, examination (MMSE) [10].
- Performance within the range of normal in the memory test. Greater scale score of 6 in all indexes of free recall test and selectively provided Buschke (FCSRT) [11].

- Autonomy in activities of daily living. Less than 6 in the functional activities questionnaire Pfeffer [12] score.

*2) MCI group*

- Men and women aged between 39 and 79 years (inclusive) to sign the informed consent to participate in the PGA.
- The presence of memory complaints that the assessor considers relevant.
- Global cognitive function preserved. Highest score of 24 in the MMSE.
- Performance below the average adjusted by age and educational level in at least one of the tests that assess memory, language, visual-spatial function, attention and executive function. Scale score equal to or less than 6.
- Autonomy in activities of daily living. Less than 6 score on the Questionnaire of Pfeffer Functional Activities

Exclusion criteria are the presence of dementia (DSM-IV and stage CDR $\geq 1$); significant history of neurological disease of any kind that might cause cognitive impairment dementia; history of diagnosed psychiatric illness that could cause dementia cognitive impairment; people with mild psychiatric symptoms (anxiety, depression) are not excluded. Finally a balanced database of about 100 individuals has been created PGA-OREKA.

### B. Task of Categorical Verbal Fluency (CVF)

The task of categorical verbal fluency (CVF) is a test to measure and quantify the progression of cognitive impairment in neurodegenerative diseases. It is widely used to assess language skills, associative memory and executive functions. During the CVF task the interviewer ask the patient to list all the names you know of a category in one minute (animals). In clinical practice, only the total number of elements emitted is recorded. However, you can get more specific indexes that detect subtle changes in cognitive status of the patient and thus gain in specificity to discriminate between different degenerative diseases [1,9].

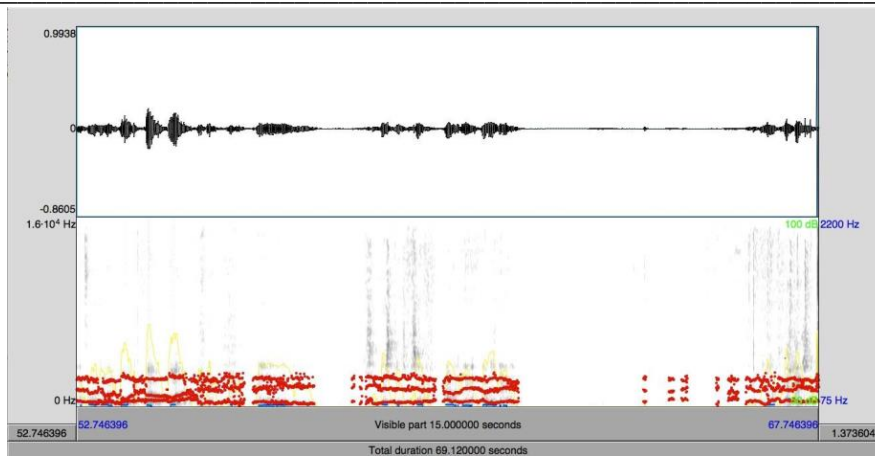

**Figure 2.** Categorical Verbal Fluency (CVF) task uttered by an individual with MCI

## III. METHODS

The analysis by automated methods of this task, possibly combined with other methodologies, could be a useful non-invasive method for early AD diagnosis [8]. The analysis of CVF task is based on three families of features (SSF set), obtained by the Praat software package [13] and software that we ourselves developed in MATLAB. For that purpose, an automatic Voice Activity Detector has extracted voiced/unvoiced segments as parts of an acoustic signal [8].

The families of features include in this experimentation are [8]:

### A. Feature extraction

#### 1) Linear features
- *Duration:* the histogram calculated over the most relevant voiced and unvoiced segments, the average of the most relevant voiced/unvoiced, voiced/unvoiced percentage and spontaneous speech evolution along the time dimension, and the voiced and unvoiced segments' mean, max and min;
- *Time domain:* short time energy. Energy, derivative of the energy.
- *Frequency domain, quality*: spectral centroid. If m is too small (smaller than 3) the algorithm will work wrongly because it will only have few distinct states for recording but it depends on the data. When using long signals, a large value of m is preferable but it would require a larger computational time.
- 12 *Mel Frecuency Cepstral Coeficients* MFCC
- *Acoustic features*: pitch, standard deviation pitch, max and min pitch, intensity, standard deviation intensity, max and min intensity, period mean, period standard deviation, and Root Mean Square amplitude (RMS).
- *Voice quality features*: shimmer, local jitter, Noise-to-Harmonics Ratio (NHR), Harmonics-to-Noise Ratio (HNR) and autocorrelation.
- *Duration features*: fraction of locally unvoiced frames, degree of voice breaks.

#### 2) Non linear features
- Castiglioni fractal fimension
- Permutation entropy

### B. Automatic selection of features by ANOVA

Then we will select automatically the best argument with regard to common significance level. Thus automatic feature selection is performed by ANOVA [14]. This function performs balanced one-way ANOVA for comparing the means of two or more columns of data in the matrix X, where each column represents an independent sample containing mutually independent observations. The function returns the p-value under the null hypothesis that all samples in X are drawn from populations with the same mean. If p is near zero, it casts doubt on the null hypothesis and suggests that at least one sample mean is significantly different than the other sample means. Common significance levels are 0.05 or 0.01 in our case, 0.05.

### C. Automatic classification

The WEKA software suite [15] has been used in carrying out the experiments. Support Vector Machines have been used for the automatic classification. The results were evaluated using Classification Error Rate (CER). For the training and validation steps, we used k-fold cross-validation with k=10. Cross-validation is a robust validation method for variable selection [16]. Repeated cross-validation (as calculated by the WEKA environment) allows robust statistical tests. We also use the measurement provided automatically by WEKA "Coverage of cases" (0.95 level) and Confidence Interval for percentages (CI) for 95%, 90% and 80%.

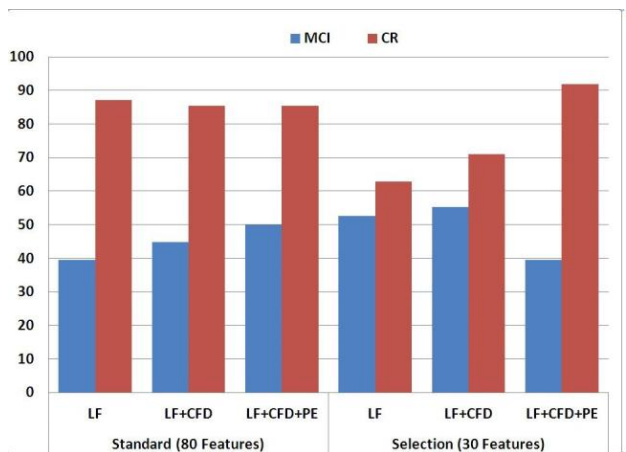

**Figure 1.** CER (%) for CR and MCI groups with SVM for standard features and selection of features by ANOVA.

IV. RESULTS AND DISCUSSION

The experimentation has been carried out with the balanced subset PGA-OREKA. This preliminary experimentation is divided in two stages.

In the first stage the linear and non-linear features described in Section III.A are used, about 80. Figure 1shows the results in the left size.

- SVM with Linear Features (LF) obtain good results for CR but the results for MCI are about 40%.
- The inclusion of Castiglioni fractal dimensions improves the results for MCI.
- LF+CFD+PE looks as the best option for both CR and MCI and in this last case the rate achieves is about %50.

In the next stage a selection of features by ANOVA is carried out. The size set is reduced more than 50%. Figure 1shows the results in the right size.

- The global results are improved in all cases.
- The best global option is LF+CFD+PE. In this case the results for CR are about 90% but for MCI decreases to 40%

V. CONCLUSION

This work, on selection of biomarkers from speech, is part of a wide-ranging cross study for the diagnosis of Alzheimer. Specifically in this work a task for detection of MCI has been used. The task analyzes Categorical Verbal Fluency. The automatic classification is carried out by SVM. Classical linear features, Castiglioni fractal dimension and Permutation Entropy are analyzed. Finally the most relevant features are selected by ANOVA test. The promising results are over 50% for MCI. In ongoing works new non-linear features, other entropy algorithms and semantic indexes will be used.

*Acknowledgments*

This work has been partially supported by the University of the Basque Country by UPV/EHU—58/14 project, SAIOTEK from the Basque Government, and the Spanish Ministerio de Ciencia e Innovación TEC2012-38630-C04-03.